\documentclass[11pt,a4paper]{article}
\pdfoutput=1
\usepackage{jheppub}
\usepackage{amsfonts}
\usepackage{bbm}
\usepackage{graphicx}
\usepackage{caption}
\usepackage{subcaption}
\usepackage{bigints}
\usepackage[normalem]{ulem}
\usepackage{slashed}

 \def\g{\gamma}

\def\e{\epsilon}

\def\m{\mu}
\def\n{\nu}

\newcommand{\RN}[1]{%
  \textup{\uppercase\expandafter{\romannumeral#1}}%
}

\newcommand{\ndt}{\noindent}

\newcommand{\nn}{\nonumber}

\def\e{\epsilon}

\def\p{\partial}

\def\bea{\begin{eqnarray}}
\def\eea{\end{eqnarray}}
\def\be{\begin{equation}}
\def\ee{\end{equation}}
\def\ba{\begin{align}}
\def\ea{\end{align}}

\newcommand{\bem}{\begin{pmatrix}}
\newcommand{\eem}{\end{pmatrix}}

\def\={\;  = \;}
\def\+{\, + \,}

\def\rt2{\sqrt{2}}

\title{Quench Disorder and Scalar Field Theory in the Presence of Boundary}

\author{\small{Rajesh Kumar Gupta}}
\affiliation{\small{Department of Physics, Indian Institute of Technology Ropar,
Rupnagar, Punjab 140001, India}}

\emailAdd{rajesh.gupta@iitrpr.ac.in}

\abstract{ Disordered systems are interesting for many physical reasons. In this article, we study the renormalization group property of quenched disorder systems in the presence of a boundary. We construct examples of scalar field theories in various dimensions with both classical and quantum disorder localized at the boundary. We study these theories in $\e$-expansion and discuss properties of fixed points of the renormalization group flow. 
}

\begin{document}

\maketitle

\section{Introduction \label{sec:intro}}
Disorder systems are of many physical interests. They are relevant in statistical physics, condensed matter and high energy physics. The physical reason for considering disorder is that, in practice, real systems are not pure, and they come with either impurities or inhomogeneous background fields. 
The presence of impurities or non-constant background fields affects the microscopic interactions of the pure system. As a result, it is clearly of interest to understand the effect of impurities on the large distance properties of the system, particularly the critical behaviour of the system. 

In many situations, the impurities are frozen in time, i.e. these can be treated as non-dynamical. These kinds of disorders are called quench disorders. Assuming that the scale of variation of the disorders is much smaller than the size of the system, we can treat them as a field varying independently and randomly at every point in space and taking values from a probability distribution. 
A given profile of a disorder breaks the translational invariance of the pure system. 
The Hamiltonian of the system is no longer homogeneous and involves inhomogeneous coupling constants that couple to one or more local operators.
In this paper, we will focus on disorders of this type. Such disorder systems have been studied previously with different motivation~\cite{Harris_1974, PhysRevB.26.154, Dotsenko:1994im, Dotsenko:1994sy, Fujita:2008rs, Hartnoll:2014cua, Hartnoll:2015rza, Aharony:2015aea, Aharony:2018mjm, Narovlansky:2018muj}

A quench disorder comes in two flavours. A $d$-dimensional classical statistical mechanical system near a second-order phase transition is described by a $d$-dimensional Euclidean field theory. A quench disorder, in this case, is called a classical disorder. The classical disorder field $h(x)$ is function of spatial coordinates and takes values from a probability distribution $P[h]$. On the other hand, a disorder in a quantum mechanical system, which has an extra time direction, is called the quantum disorder. In this case the disorder field $h(x)$ is only function of spatial coordinates and takes values from a probability distribution $P[h]$. We will analytically continue the time direction and work with a Euclidean theory to treat classical and quantum disorders uniformly. In that case, the quantum disorder field $h(x)$ will be independent of the Euclidean time coordinate. 

A disordered system is specified by the knowledge of various thermodynamical quantities and correlation functions of local operators. These properties of the system depend on the disorder profile $h(x)$, which takes values from the probability distribution $P[h]$. Often, most notably near the critical point, it is not sufficient to know the property of the system for a given realization of the disorder. To characterize the disorder system, one needs an ensemble of many different systems with the different realizations of the disorder profile (with the given probability distribution). As a result, the free energy and thermodynamic quantities obtained by differentiating the free energy are averaged over the probability distribution. 

The present paper aims to compute the disordered averaged correlation function of local operators, and critical exponents near the critical point of the renormalization group flow. These are relevant in the context of the second-order phase transition in the disordered system.
It is important to emphasize that typically, even a small amount of disorder can lead to a significant change in the critical behaviour of the system. A disorder perturbation can grow at long distances leading to a nontrivial fixed point or no nontrivial fixed point.

In this paper, we study the effect of a random quench disorder on a conformal field theory. More precisely, we begin with a pure system described by a conformal field theory and deform it by one or more interactions. For simplicity, we will consider one of the interactions to be a disorder interaction that couples to a scalar operator and leads to a renormalization group flow in the space of coupling constant to a new fixed point (or no fixed point).
 The conformal field theories of our interests will be boundary conformal field theories (bCFTs). Specifically, we will consider a special class of boundary conformal field theories obtained by a free scalar field theory in $(d+1)$-dimensional Euclidean space. These classes of bCFTs have been studied before in many different contexts with various motivations; see for example~\cite{Callan:1993mw, Callan:1994ub, Fendley:1994rh, Herzog:2017xha, Giombi:2019enr}. We will investigate the critical properties of these bCFTs deformed by disorder interaction localized at the $d$-dimensional planar boundary. Our computations will be in the framework of $\e$-expansion. We will find examples where, to leading order in $\e$, weakly coupled disorder Wilson-Fisher fixed point exists together with weakly coupled pure Wilson-Fisher fixed point. These provide examples of field theories with IR disorder fixed points. Another property of this class of bCFTs, is the existence of an infinite number of higher spin currents. These currents are conserved in bulk with the partial breaking of conservation law at the boundary, i.e., a delta function source localized at the boundary for the divergence of these currents. These sources define boundary operators, which are called higher spin displacement operators. These operators are protected, i.e. they do not receive anomalous dimension even though there are interactions localized at the boundary. We find that the statement is true even in the presence of disorder interactions.

The organization of this paper is as follows: In section 2, we introduce the replica treatment of quench disorder in field theory in the presence of a boundary. We then discuss the renormalization group flow and the generalization of the Callan-Symanzik equation in disordered quantum field theory. In section 3, we briefly discuss the salient features of a free scalar field theory in the presence of a planar boundary. In section 4, we study a few examples of free bulk scalar field theory in the presence of a planar boundary with Neumann boundary condition. The scalar field is  interacting with boundary degrees of freedom in the presence of classical disorder interactions localized at the boundary. We find the fixed point of the boundary renormalization group flow, and calculate the anomalous dimension of $(\text{mass})^{2}$ operator at the fixed point. Section 5 discusses the anomalous dimension of the higher spin displacement operator localized at the boundary. In section 6, we study an example of field theory in the presence of quantum disorder localized at the boundary and look for the quantum disorder fixed point. Finally, in section 7, we conclude with a brief discussion.  
\section{Disorder in quantum field theory, replica trick and renormalization group flow}
This section will discuss the quenched disorder in Euclidean field theory in the presence of a planar boundary. Our discussion follows very closely the work of~\cite{Aharony:2018mjm, Narovlansky:2018muj} where the authors examined the renormalization group flow in Euclidean field theory (without boundary) in the presence of quench disorder. We will apply their work in our set up where we will have a Euclidean field theory in the presence of a boundary with a disorder interaction localized at the boundary. The analysis in the present section uses the replica trick, which is a very practical approach to dealing with quench disorder.  

Our conventions are as follows. We will take $d$-number of spatial coordinates denoted by $\vec x$ to be the boundary coordinates and $y\geq 0$ is the direction normal to the boundary, i.e. the bulk is labelled by $x\equiv (\vec x, y)$ and the boundary is at $y=0$. Thus, for the discussion of the classical disorder, the bulk dimension is $(d+1)$ and the boundary dimension is $d$. When we discuss a quantum disorder, we also need to introduce an extra time coordinate $t$. In this case, the bulk coordinates are $(t,\vec x, y)$ and the boundary coordinates are $(t,\vec x)$. Accordingly, the bulk and boundary dimensions are $(d+2)$ and $(d+1)$, respectively. A quench disorder localized at the boundary will be denoted by $h(\vec x)$, i.e. the coupling constant varies as a function of spatial coordinates $\vec x$. In particular, in the case of a quantum disorder, the disorder field is independent of the time coordinate and varies only spatially.  

We will further make the disorder random with a probability distribution that we denote by $P[h]$. We will assume that probability distribution given by a Gaussian function,
\be
P[h]=\mathcal N\exp\Big(-\frac{1}{2v}\int d^{d}\vec x\,h^{2}(\vec x)\Big)\,,
\ee 
where $\mathcal N$ is a constant determined by the normalization condition
\be
\int[Dh]P[h]=1\,.
\ee
For the Gaussian distribution, we have
\be
\overline{h(\vec x)}=0,\quad \overline{h(\vec x)h(\vec x')}=v\,\delta^{d}(\vec x-\vec x')\,,
\ee
and higher moments are obtained by Wick's contraction.

Now, consider a (pure) Euclidean field theory in the presence of boundary described by an action $S_{0}$. The action $S_{0}$ will consist of bulk fields interacting with degrees of freedom localized at the boundary. For example, our case of interests will be where the bulk fields are free scalar fields having nontrivial interactions with degrees of freedom propagating on a boundary.
We introduce the disorder at the boundary by coupling the field $h(\vec x)$ to a boundary operator $\mathcal O_{0}(\vec x)$. Assuming that the operator is a scalar operator, the action for the classical disordered theory is 
\be
S=S_{0}+\int d^{d}\vec x\,h(\vec x)\mathcal O_{0}(\vec x)\,.
\ee
In the case of the quantum disorder the corresponding action is
\be
S=S_{0}+\int d^{d}\vec x\,dt\,h(\vec x)\mathcal O_{0}(t,\vec x)\,.
\ee
Note that one can consider a more general situation where there are more than one disordered coupling constant. Next, we want to compute the correlation function of operators in a disordered theory.

We will begin with the case of classical disorder. The partition function for a given profile of the disorder is
\be
Z[h,J_{i}]=e^{W[h,J_{i}]}=\int\mathcal D\phi\,e^{-S-\int d^{d}\vec x\,h(\vec x)\mathcal O_{0}(\vec x)+\sum_{i}\int d^{d+1}x\,J_{i}\mathcal O_{i}}\,.
\ee 
Here $W[h,J_{i}]$ is the generating functional for the connected correlation function for a given profile of the disorder field~\footnote{Here, we have turned on the sources for the bulk operators. It is just for the presentation. We could also turn on sources for the operators localized on the boundary.}. We are interested in the disorder averaged connected correlation function. The generating functional for the disordered average connected correlation function is given by
\be\label{DisW}
W_{D}[J_{i}]=\int [Dh]P[h]W[h,J_{i}]\,,
\ee
and the disordered averaged connected correlation functions are 
\bea
\overline{<\mathcal O_{1}(x_{1})\mathcal O_{2}(x_{2})....\mathcal O_{n}(x_{n})>_{\text{conn.}}}&=&\int \mathcal Dh\,P[h]\,<\mathcal O_{1}(x_{1})\mathcal O_{2}(x_{2})....\mathcal O_{n}(x_{n})>_{h,\text{conn.}}\,,\nn\\
&=&\frac{\delta^{n}W_{D}[J_{i}]}{\delta J_{1}(x_{1})....\delta J_{n}(x_{n})}\Big|_{J_{i}=0}\,.
\eea 
We calculate the disordered average free energy using the replica trick. In the replica approach, we compute the partition function of $n$-copies of the original theory maintaining the replica symmetry i.e.
\be
Z^{n}[h,J]=\int \prod_{A=1}^{n}D\phi_{A}\,e^{-\sum_{A=1}^{n}S_{A}-\sum_{A=1}^{n}\int d^{d}\vec x\,h(\vec x)\mathcal O_{0,A}(\vec x)+\sum_{i,A}\int d^{d+1}x\,J_{i}\mathcal O_{i,A}}\,.
\ee
Next, we introduce the replica functional
\bea
W_{n}[J]&=&\int\, [Dh]P[h]Z^{n}[h,J]\,,\nn\\
&=&\int\prod_{A=1}^{n}D\phi_{A}\,e^{-S_{\text{repl.}}+\sum_{i}\sum_{A}\int d^{d+1}x\,J_{i}\mathcal O_{i,A}}\,,
\eea
where $S_{\text{repl.}}$ is the replica action obtained after integrating over the disorder field and is given by
\be
S_{\text{repl.}}=\sum_{A=1}^{n}S_{A}-\frac{v}{2}\sum_{A,B=1}^{n}\int\,d^{d}\vec x\,\mathcal O_{0,A}(\vec x)\mathcal O_{0,B}(\vec x)\,.
\ee
Note that the replicated action has translation and rotation invariance.  
The disordered free energy is then obtained as
\be\label{WD.1}
W_{D}[J_{i}]=\lim_{n\rightarrow 0}\frac{\p W_{n}[J]}{\p n}\,.
\ee
A generalization of the above discussion to the case of quantum disorder is straightforward. However, there are some modifications to the replica action, which is essential to emphasize. As we stated above, the quantum system is defined on the $(d+1)$-dimensional space where fields are a function of $(d+1)$-spatial coordinates and a time coordinate. The disorder field $h(\vec x)$ at the boundary is the only function of the $d$-spatial coordinates. Therefore, for a specific profile of the disorder field the translation and rotation invariances of the boundary are broken but the time translational invariance is maintained. In this case, the partition function is 
\be
Z[h,J_{i}]=e^{W[h,J_{i}]}=\int\mathcal D\phi\,e^{-S-\int dt\,d^{d}\vec x\,h(\vec x)\mathcal O_{0}(t,\vec x)+\sum_{i}\int dt\,d^{d}x\,dy\,J_{i}\mathcal O_{i}(t,\vec x,y)}\,,
\ee 
and the replica functional is
\bea
W_{n}[J]&=&\int \prod_{A=1}^{n}D\phi_{A}\,e^{-S_{\text{repl.}}+\sum_{i,A}\int dt\,d^{d}x\,dy\,J_{i}\mathcal O_{i,A}(t,\vec x,y)}\,,
\eea
where $S_{\text{repl.}}$ is the replica action obtained after integrating over the disorder field and is given by
\be
S_{\text{repl.}}=\sum_{A=1}^{n}S_{A}-\frac{v}{2}\sum_{A,B=1}^{n}\int\,dt\,dt'\,d^{d}\vec x\,\mathcal O_{0,A}(t,\vec x)\mathcal O_{0,B}(t',\vec x)\,.
\ee
We see from the replica action that it is non-local in time. As we will see later, this fact brings an extra complication of infrared divergence in our computations. Since we will be interested in calculating fixed points in the renormalization group flow, we will ignore the issues related to infrared divergences.

The generating functional for the quantum disordered correlation function is
\be\label{WD.2}
W_{D}[J_{i}]=\lim_{n\rightarrow 0}\frac{\p W_{n}[J]}{\p n}\,.
\ee
Unlike the classical case, the averaging over the quantum disorder does not restore the full invariance of the original (pure) action. 

 Using the relations~\eqref{WD.1} and~\eqref{WD.2}, we see that the disordered connected correlation function is given in terms of correlation functions in the replicated theory as
 \be\label{DisorderConnCorr.1}
 \overline{<\mathcal O_{1}(x_{1})\mathcal O_{2}(x_{2})....\mathcal O_{n}(x_{n})>_{\text{conn.}}}=\lim_{n\rightarrow 0}\frac{\p}{\p n}<\sum_{A_{1}}\mathcal O_{1,A_{1}}(x_{1})....\sum_{A_{n}}\mathcal O_{n,A_{n}}(x_{n})>_{\text{rep.}}\,.
 \ee
Note that the correlation function that appears on the right hand side is not necessarily a connected one. We can further simplify the right hand side using the replica symmetry. Assuming that the replica symmetry is not spontaneously broken, we have
\be
<\sum_{A_{1}}\mathcal O_{1,A_{1}}(x_{1})....\sum_{A_{n}}\mathcal O_{n,A_{n}}(x_{n})>_{\text{rep.}}=n<\mathcal O_{1,1}(x_{1})....\sum_{A_{n}}\mathcal O_{n,A_{n}}(x_{n})>_{\text{rep.}}\,.
\ee
Thus, we get
\be
 \overline{<\mathcal O_{1}(x_{1})\mathcal O_{2}(x_{2})....\mathcal O_{n}(x_{n})>_{\text{conn.}}}=\lim_{n\rightarrow 0}<\mathcal O_{1,1}(x_{1})\sum_{A_{2}}\mathcal O_{2,A_{2}}(x_{2})....\sum_{A_{n}}\mathcal O_{n,A_{n}}(x_{n})>_{\text{rep.}}\,.
 \ee
 
 \ndt{\bf Harris criteria:}
Thinking of the disorder as an interaction in perturbation theory, it is important to know when the interaction is relevant at long distance. It is clear that it should depend on the dimension of the operator $\mathcal O_{0}(x)$ the disorder couples to. Harris criteria~\cite{Harris_1974} provides a useful way to organize the disorder interactions in terms of relevant, irrelevant and marginal perturbations. Let us suppose the dimension of the operator $\mathcal O_{0}(x)$ is $\Delta_{0}$. From the Gaussian distribution, the disorder has dimension $[v]=2[h]-d$. Now, for the classical disorder the dimension of $h$ is $[h]=d-\Delta_{0}$. Then the classical disorder is relevant if $\Delta_{0}<\frac{d}{2}$, irrelavant for $\Delta_{0}>\frac{d}{2}$ and marginal for $\Delta_{0}=\frac{d}{2}$. For the quantum disorder, we have a slightly different condition. In this case, the dimension of the disorder field is $[h]=d+1-\Delta_{0}$. Therefore, the quantum disorder is relevant if $\Delta_{0}<\frac{d+2}{2}$, marginal for $\Delta_{0}=\frac{d+2}{2}$ and irrelevant for $\Delta_{0}>\frac{d+2}{2}$.

\ndt{\bf Renormalization group equation:}
In a standard quantum field theory, where coupling constants are homogeneous in spacetime, a renormalized correlation function satisfies Callan-Symanzik equation. A generalization of the Callan-Symanzik equation exists for the disordered case. In the case of a classical disorder, the replicated action describes a standard quantum field theory. The replicated action has an extra coupling constant compared to pure theory which is proportional to the disorder strength. Therefore, we expect that a $k$-point renormalized correlation function in the replicated theory to satisfy
\bea
&&\Big(\m\frac{\p}{\p\m}+\sum_{i}\beta_{\lambda_{i}}\frac{\p}{\p\lambda_{i}}+\beta_{v}\frac{\p}{\p v}\Big)<\mathcal O_{A_{1}}(x_{1})\mathcal O_{A_{2}}(x_{2})...>_{\text{rep.}}\nn\\
&&\qquad\qquad\qquad\qquad\qquad\qquad\,+\sum_{B_{1}}\g_{A_{1}B_{1}}<\mathcal O_{B_{1}}(x_{1})\mathcal O_{A_{2}}(x_{2})...>_{\text{rep.}}+...=0\,.
\eea
Here $A_{i}$'s and $B_{i}$'s are replica index, $\m$ is the renormalization scale and $\lambda_{i}$'s are coupling constants of the pure theory. Note that there is a beta function, $\beta_{v}$, for the disorder strength. Also, in the above equation, we have assumed that $\mathcal O_{A}(x)$ are lowest dimension operators in which case these operators mix among themselves. As a result, the anomalous dimension has the form
\be
\g_{AB}=\g\,\delta_{AB}+\g'\,,
\ee
where $\g'$ is a constant and independent of replica index.
Using the relation~\eqref{DisorderConnCorr.1}, we obtain the Callan-Symanzik equation satisfied by the $k$-point disordered average connected correlation function
\be
\Big(\m\frac{\p}{\p\m}+\sum_{i}\beta_{\lambda_{i}}\small|_{n=0}\frac{\p}{\p\lambda_{i}}+\beta_{v}\small|_{n=0}\frac{\p}{\p v}+k\,\g\small|_{n=0}\Big)\overline{<\mathcal O(x_{1})\mathcal O(x_{2})...>_{\text{conn.}}}=0\,.
\ee
The above equation describes the beta and gamma functions for the disordered theory.

Similar Callan-Symanzik equation exists for the quantum disorder case. However, the Callan-Symanzik equation for the case with quantum disorder differs from the classical disorder.  Note that the presence of quantum disorder breaks the isotropy between space and time. As a result, even though we start with a pure theory, where space and time have the same scaling, the disordered theory at the quantum critical point may not respect the scaling symmetry of space and time; for example the correlation function of the boundary operators may not respect the scaling symmetry of the boundary coordinates $(\vec x,t)\rightarrow (\lambda\,\vec x,\lambda\, t)$. In this case, the Lifshitz scaling may emerge. It was shown in~\cite{Aharony:2018mjm,Narovlansky:2018muj} that the disordered averaged $k$-point connect correlation function satisfies
\be
\Big(\m\frac{\p}{\p\m}+\sum_{i}\beta_{\lambda_{i}}\Big|_{n=0}\frac{\p}{\p\lambda_{i}}+\beta_{v}\Big|_{n=0}\frac{\p}{\p v}+\g_{t}\sum_{i=1}^{k}t_{i}\frac{\p}{\p t_{i}}+k\,\g\Big|_{n=0}\Big)\overline{<\mathcal O(x_{1})\mathcal O(x_{2})...>_{\text{conn.}}}=0\,.
\ee
The critical exponent $\g_{t}$ gives the Lifshitz scaling exponent
\be
z=1+\g_{t}^{*}\,,
\ee
with $\g_{t}^{*}$ evaluated at the quantum disorder fixed point.
\section{Scalar field theory in the presence of boundary}
Our goal in this article is to study the renormalization group flow in a scalar field theory when one of the interactions at the boundary is a disorder interaction. More specifically, we will focus on a class of models where bulk interactions have been switched off i.e. the scalar fields in the bulk are free fields and having interactions localized at the boundary. For example in the case of classical disorder, the general action of our interests would be 
\be
S=\int d^{d}\vec x\,dy\,\frac{1}{2}\p_{\m}\phi^{I}\p^{\m}\phi^{I}+\int d^{d}\vec x\,\mathcal L_{\text{int}}+\int d^{d}\vec x\,\mathcal L_{\text{dis.int}}\,,
\ee
where $I=1,...,N$, and $\mathcal L_{\text{int}}$ and $\mathcal L_{\text{dis.int}}$ are relevant/marginal and disorder interactions, respectively. Furthermore, because of the presence of boundary, the scalar fields need to satisfy either Dirichlet or generalized Neumann condition. In this paper, we will be focussing on examples with generalized Neumann boundary condition.

A free scalar field theory is a conformal field theory. In the dimensions, $d+1$, the conformal group is $SO(d+2,1)$. The presence of co-dimension one planar boundary breaks the conformal symmetry to the conformal group of the boundary, i.e. $SO(d+1,1)$. The energy-momentum tensor is not conserved in all directions, and the divergence of the energy-momentum tensor along the direction perpendicular to the boundary defines a displacement operator,
\be\label{DisplacementOp}
\p_{\m}T^{\m y}=D(\vec x)\delta(y)\,.
\ee
Since the displacement operator is obtained from the energy-momentum tensor, which is conserved in the bulk, one would expect it to be a protected operator, i.e. the scaling dimension of the displacement operator is the same as the energy-momentum tensor. We will see that this is true in the presence of disorder interactions at the boundary.

A free field theory has also an infinite number of higher spin conserved currents. These currents are bilinear in fields and constructed out by acting derivatives on them. For example in the case of a free scalar field theory, a spin $s$-current is given by~\cite{Giombi:2016ejx}
\be
J^{s}_{\m_{1}...\m_{s}}=\sum_{k=0}^{s}a_{sk}\,\p_{\{\m_{1}...\m_{k}}\phi^{I}\,\p_{\m_{k+1}...\m_{s}\}}\phi^{I}\,.
\ee
Here $a_{sk}$'s are constants determined by the conservation condition, and curly brackets ensure traceless symmetrization condition. The dimension of a spin $s$-current is $\Delta_{s}=d-1+s$. 
The presence of interactions localized at the boundary has an interesting consequence on the conservation of higher spin currents. These currents are conserved in bulk, where fields are free; however, the conservation law is broken at the boundary.
Following the definition of the displacement operator~\eqref{DisplacementOp}, one can define the higher spin analog of the displacement operator~\cite{Giombi:2019enr}.
More precisely, we define
\be
\p^{\m}J^{s}_{\m\m_{1}...\m_{s-2}y}=D_{\m_{1}..\m_{s-2}}(\vec x)\delta(y)\,.
\ee
Note that the boundary operator $D_{\m_{1}..\m_{s-2}}(\vec x)$ carries bulk index $\m=(i,y)$, where $i$ is the boundary index. When all $\m$'s equal to $y$, we obtain a spin zero boundary operator whereas all $\m$'s different from $y$ corresponds to a spin $(s-2)$-operator. Thus, the boundary operator $D_{\m_{1}..\m_{s-2}}(\vec x)$ gives rise operators of all spins between $0$ and $s-2$. The classical dimensions of these boundary operators are the same as the dimension of the current, i.e. $\Delta=d-1+s$. One would expect that the scaling dimension of the higher spin displacement operators will not renormalize in the perturbation theory. As we will see later, the higher spin displacement operator does not have an anomalous dimension due to disorder interaction at the boundary.

\section{Classical disorder at boundary}
This section discusses examples of free scalar field theories with classical disorder interaction turned on on the boundary. We will restrict ourselves to the cases where the disorder interaction is marginal. We will then look for the fixed point of the renormalization group flow.
\subsection{Single scalar field with a classical disorder at boundary}
We start with a simplest example of a scalar field theory with a classical disorder interaction at the boundary. The action is given by
\be\label{SingleScalar}
S=\int d^{d}\vec x\,dy\,\frac{1}{2}\p_{\m}\phi\p^{\m}\phi+\int d^{d}\vec x\,h(x)\phi^{2}(x)+\frac{\lambda}{4!}\int d^{d}\vec x\,\phi^{4}(x)\,.
\ee
The strength of the disorder is marginal in the dimension $d=2$ and is relevant for $d<2$. We have also added $\phi^{4}$ interaction since it is a marginal in $d=2$ dimensions.
We will therefore study the above theory in the dimensions $d=2-\e$. 
The action for the replicated theory is given by
\bea\label{ReplicatedSingleScalar}
S_{\text{repl.}}&=&\sum_{A=1}^{n}S_{A}-\frac{v}{2}\int d^{d}x\,\sum_{A,B=1}^{n}\phi_{A}^{2}(x)\mathcal \phi^{2}_{B}(x)\,,\nn\\
&=&\int d^{d}x\,dy\,\frac{1}{2}\sum_{A=1}^{n}\p_{\m}\phi_{A}\p^{\m}\phi_{A}+\frac{\lambda}{4!}\sum_{A=1}^{n}\int d^{d}x\,\phi_{A}^{4}-\frac{v}{2}\int d^{d}x\,\sum_{A,B=1}^{n}\phi_{A}^{2}(x)\mathcal \phi^{2}_{B}(x)\,.\nn\\
\eea
Thus, the replicated theory is a standard quantum field theory and we will treat the disorder in a perturbation theory.

\begin{figure}[htpb]
\begin{center}
\includegraphics[width=6in]{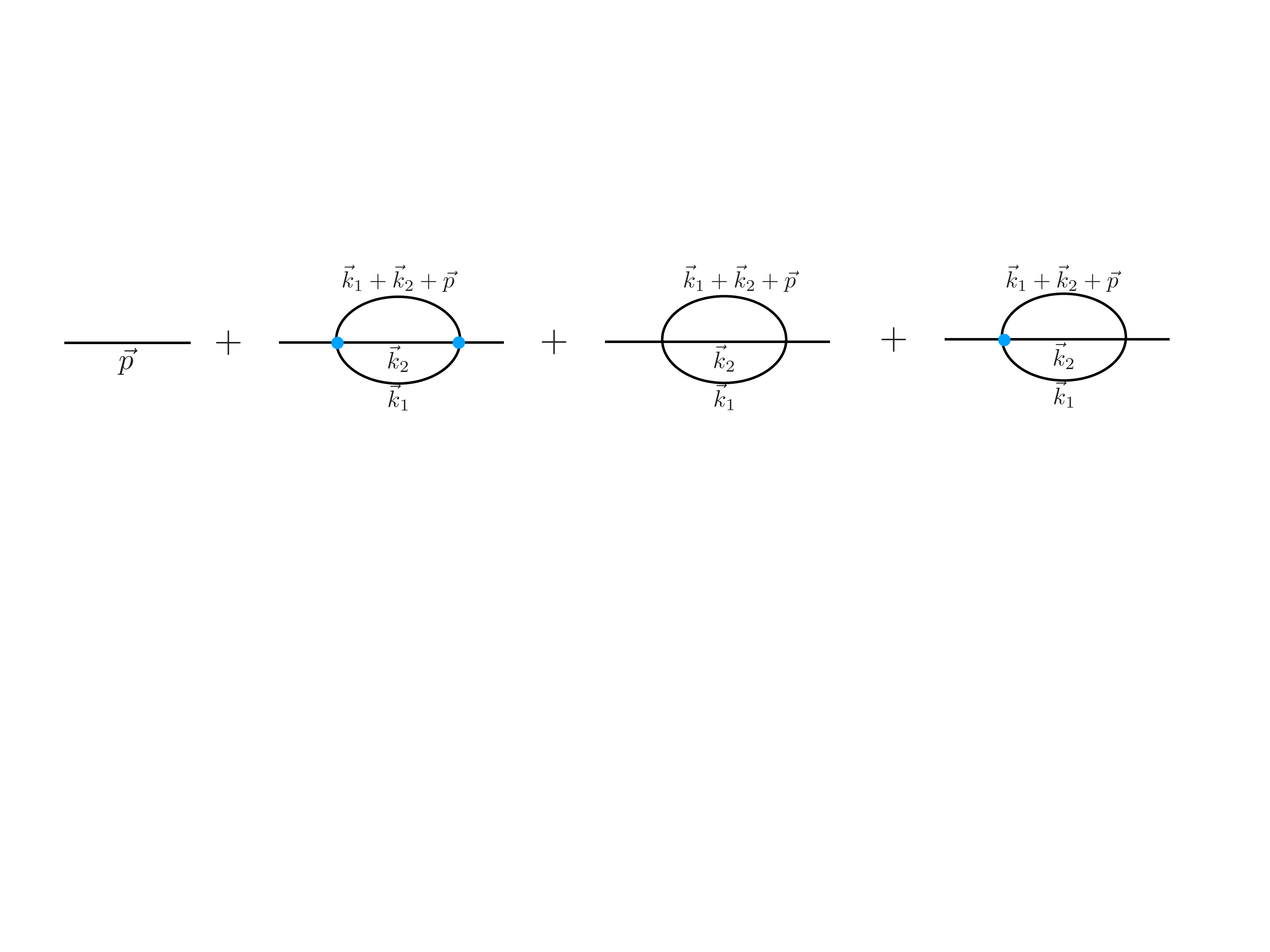}
\vspace{-4cm}
\caption{Bulk 2-point function at 2-loop order. The blue dot represents the disorder vertex.\label{2-pointFn1}}
\end{center}
\end{figure}

Since there are no bulk interactions, we expect the anomalous dimension of the scalar field to be zero. We see this by noting that there is no wave function renormalization. 
We compute the bulk 2-point function $<\phi_{A}(\vec p,y_{1})\phi_{B}(-\vec p,y_{2})>$. When we fine tune the mass term to zero, the first non trivial contributions to the 2-point function comes from 2-loop diagrams, shown in figure. \ref{2-pointFn1}. 
The contributions of these Feynman diagrams are 
\bea
&&\frac{e^{-p|y_{1}-y_{2}|}+e^{-p(y_{1}+y_{2})}}{2p}\delta_{AB}+\Big(8v^{2}(n+1)+\frac{\lambda^{2}}{6}-4\lambda\,v\Big)\frac{e^{-p(y_{1}+y_{2})}}{p^{2}}\delta_{AB}\times\nn\\
&&\times\int\frac{d^{d}\vec k_{1}d^{d}\vec k_{2}}{(2\pi)^{2d}}\frac{1}{|\vec k_{1}||\vec k_{2}||\vec k_{1}+\vec k_{2}+\vec p|}\,,\nn\\
&&=\frac{e^{-p|y_{1}-y_{2}|}+e^{-p(y_{1}+y_{2})}}{2p}\delta_{AB}+\Big(8v^{2}(n+1)+\frac{\lambda^{2}}{6}-4\lambda\,v\Big)\frac{e^{-p(y_{1}+y_{2})}}{p^{2}}\delta_{AB}\frac{\Gamma(\frac{d-1}{2})^{3}\Gamma(\frac{3-d}{2})}{(4\pi)^{d}\pi^{3/2}\Gamma(\frac{3d-3}{2})(p^{2})^{\frac{3-2d}{2}}}\,.\nn\\
\eea
For $d=2-\e$, we see that there is no divergence. 
Thus, the disordered averaged connected 2-point function at the fixed point is given by
\bea
\overline{<\phi(\vec p,y_{1})\phi(-\vec p,y_{2})>_{\text{conn.}}}&=&\lim_{n\rightarrow 0}\frac{\p}{\p n}<\sum_{A}\phi_{A}(\vec p,y_{1})\sum_{B}\phi_{B}(-\vec p,y_{2})>\,,\nn\\
&=&\frac{e^{-p|y_{1}-y_{2}|}+e^{-p(y_{1}+y_{2})}}{2p}+\Big(8v_{*}^{2}+\frac{\lambda_{*}^{2}}{6}-4\lambda_{*}\,v_{*}\Big)\frac{e^{-p(y_{1}+y_{2})}}{8\pi^{2}\,p}\,.\nn\\
\eea
Here $\lambda_{*}$ and $v_{*}$ are the values of the coupling constants at the fixed point of the disordered theory.

\begin{figure}[htpb]
\begin{center}
\vspace{-0.2 cm}
\includegraphics[width=6in]{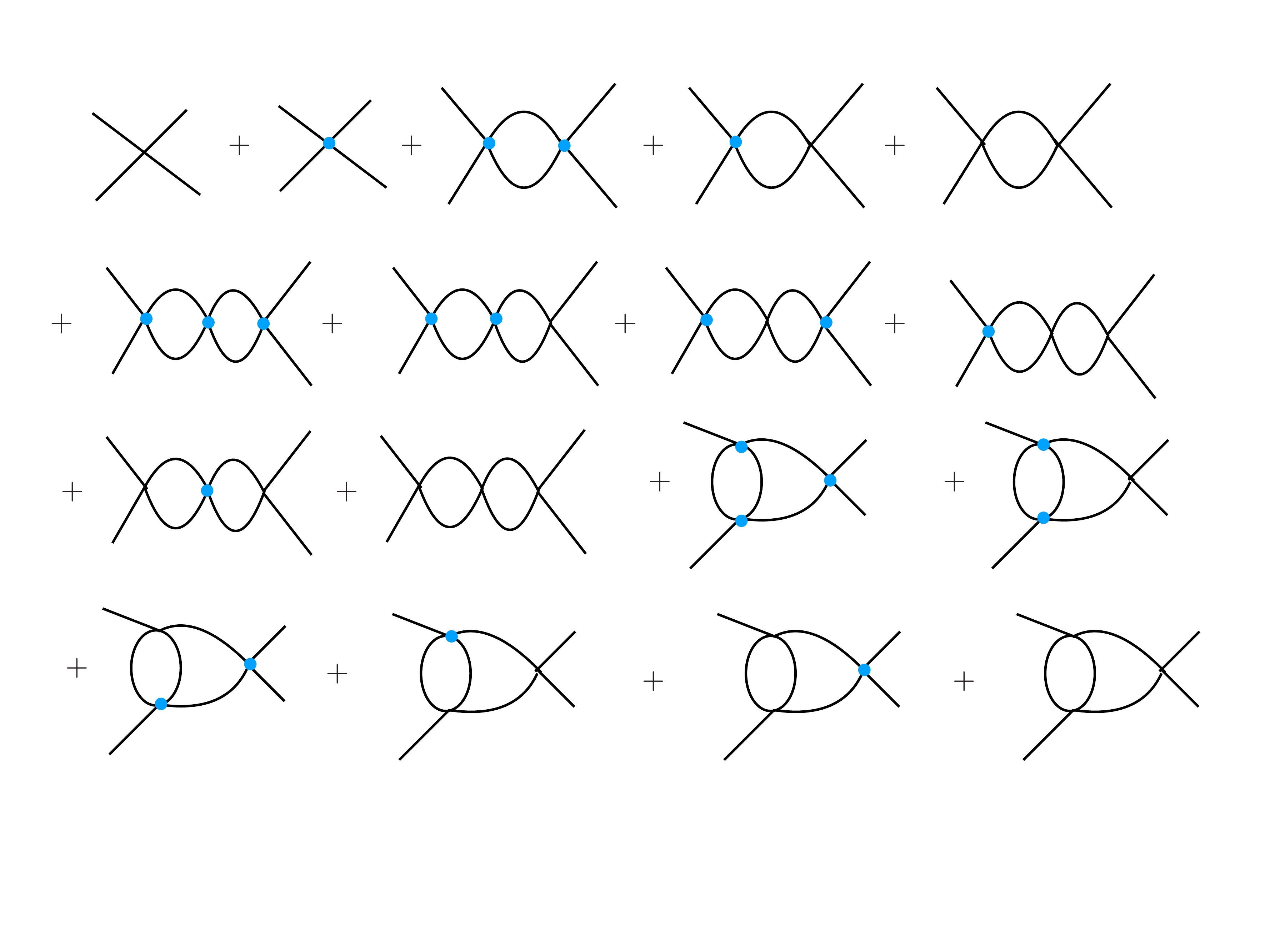}
\vspace{-2 cm}
\caption{Boundary 4-point function at 2-loop order. The blue dot represents the disorder vertex.\label{4-pointFn1}}
\end{center}
\end{figure}

\begin{figure}[htpb]
\begin{center}
\vspace{-2 cm}
\includegraphics[width=6in]{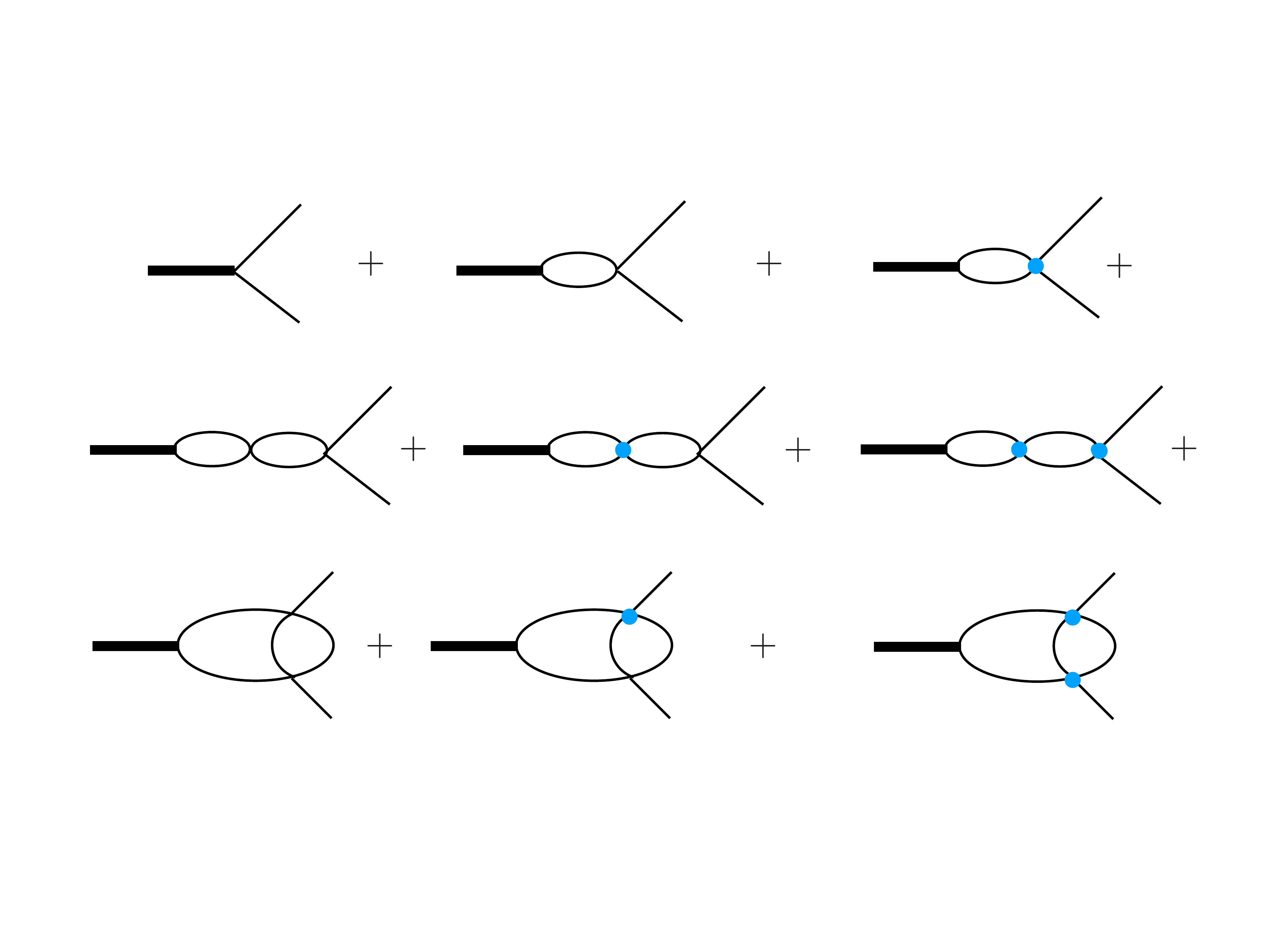}
\vspace{-2 cm}
\caption{Anomalous dimension computation for $\phi^{2}$ operator. The blue dot represents the disorder vertex.\label{CompositeOperator1}}
\end{center}
\end{figure}

Next, we compute the 4-point function of the boundary value of scalar fields. In particular, we compute $<\phi_{A}(\vec p_{1})\phi_{B}(\vec p_{2})\phi_{C}(\vec p_{3})\phi_{D}(\vec p_{4})>$ upto 2-loop order. The diagrams are shown in the figure \ref{4-pointFn1}.
The contributions are
\bea
&&-\lambda(1+\delta_{\lambda})\delta_{A,B,C,D}+4v(1+\delta_{v})\delta_{AB}\delta_{CD}+\Big(8v^{2}(1+\delta_{v})^{2}(n+8)\delta_{AB}\delta_{CD}+\frac{3}{2}\lambda^{2}(1+\delta_{\lambda})^{2}\delta_{A,B,C,D}\nn\\
&&-4\lambda v(1+\delta_{\lambda})(1+\delta_{v})(\delta_{AB}\delta_{CD}+6\delta_{A,B,C,D})\Big)I_{1}(\vec p)\nn\\
&&+\Big(16v^{3}\delta_{AB}\delta_{CD}(n^{2}+6n+20)-12v^{2}\lambda \delta_{AB}\delta_{CD}(n+4)-144v^{2}\lambda \,\delta_{A,B,C,D}+3v\lambda^{2}\delta_{AB}\delta_{CD}\nn\\
&&+18v\lambda^{2}\delta_{A,B,C,D}-\frac{3\lambda^{3}}{4}\delta_{A,B,C,D}\Big)I_{2}(\vec p)\nn\\
&&+\Big(64v^{3}(5n+22)\delta_{AB}\delta_{CD}-48v^{2}\lambda(n+14)\delta_{A,B,C,D}-192v^{2}\lambda\,\delta_{AB}\delta_{CD}+96v\lambda^{2}\delta_{A,B,C,D}\nn\\
&&+4v\lambda^{2}\delta_{AB}\delta_{CD}-3\lambda^{3}\delta_{A,B,C,D}\Big)I_{3}(\vec p,\vec q)\,.
\eea
In the above the notation $\delta_{A,B,C,D}$ is defined as
\be
\delta_{A,B,C,D}=\begin{cases} 1,\quad\text{if}\,\,A=B=C=D\,,\\
0,\quad\text{otherwise}\,.\end{cases}
\ee
The integrals $I_{1,2,3}$ are given in the appendix~\ref{UsefulIntegrations}. These integrals are divergent. Expanding near $d=2-\e$, and requiring that the 4-point function to be finite determines the counter terms $\delta_{\lambda}$ and $\delta_{v}$. These are given by
\bea
&&\delta_{\lambda}=\frac{3(\lambda-16v)}{4\pi\e}-\frac{48(n+14)v^{2}-96v\lambda+3\lambda^{2}}{4\pi^{2}\e}\ln 2+\frac{96(20+n)v^{2}-264v\lambda+9\lambda^{2}}{16\pi^{2}\e^{2}}\,,\nn\\
&&\delta_{v}=\frac{\lambda-2(n+8)v}{2\pi\e}-\frac{16(22+5n)v^{2}-48v\lambda+\lambda^{2}}{4\pi^{2}\e}\ln 2+\frac{16(8+n)^{2}v^{2}-12(n+12)v\lambda+5\lambda^{2}}{16\pi^{2}\e^{2}}\,.\nn\\
\eea
The beta function equations are given by
\bea
\e\lambda+\beta_{\lambda}+\frac{\lambda}{1+\delta_{\lambda}}\Big(\frac{\p\delta_{\lambda}}{\p{\lambda}}\beta_{\lambda}+\frac{\p\delta_{\lambda}}{\p{v}}\beta_{v}\Big)=0\,,\quad \e v+\beta_{v}+\frac{v}{1+\delta_{v}}\Big(\frac{\p\delta_{v}}{\p{\lambda}}\beta_{\lambda}+\frac{\p\delta_{v}}{\p{v}}\beta_{v}\Big)=0\,.
\eea
In the above we have included the fact that in $d=2-\e$, both $\lambda$ and $v$ carries the mass dimension $\e$, whereas $\delta_{\lambda}$ and $\delta_{v}$ are dimensionless.\\
Solving the above two, we obtain
\bea
&&\beta_{\lambda}=-\e\lambda+\frac{3\lambda(\lambda-16v)}{4\pi}-\frac{3\lambda(16(n+14)v^{2}-32v\lambda+\lambda^{2})\ln 2}{2\pi^{2}}\nn\\
&&\beta_{v}=-\e v+\frac{v(\lambda-2(n+8)v)}{2\pi}-\frac{v(16(22+5n)v^{2}-48v\lambda+\lambda^{2})\ln 2}{2\pi^{2}}
\eea
Thus, the beta functions of the disordered theory are
\bea
&&\beta_{\lambda}\Big|_{n=0}=-\e\lambda+\frac{3\lambda(\lambda-16v)}{4\pi}-\frac{3\lambda(224v^{2}-32v\lambda+\lambda^{2})\ln 2}{2\pi^{2}}\nn\\
&&\beta_{v}\Big|_{n=0}=-\e v+\frac{v(\lambda-16v)}{2\pi}-\frac{v(352v^{2}-48v\lambda+\lambda^{2})\ln 2}{2\pi^{2}}
\eea
Next, we look at the fixed points which are solutions of the equations $\beta_{\lambda}\small|_{n=0}=0$ and $\beta_{v}\small|_{n=0}=0$. We find three fixed points upto $\mathcal O(\e^{2})$:
\begin{enumerate}
\item {\bf Gaussian fixed point:} This corresponds to the fixed point $\lambda=0=v$. 
\item {\bf Pure fixed point:} This corresponds to the fixed point $\lambda=\frac{4\pi\e}{3}+\frac{32\pi\e^{2}}{9}\ln 2$ and $v=0$.
\item {\bf Disorder fixed point:} This corresponds to the fixed point $\lambda=0$ and $v=-\frac{\pi\e}{8}-\frac{11\pi\e^{2}}{32}\ln 2$. For $0<|\e|<<1$, this gives rise to a UV fixed point. Note that the fixed point disappears for $\e<-0.52$. Therefore, it is crucial to go higher-order in perturbation expansion to see if the fixed point survives.
\end{enumerate}
Next, we compute the anomalous dimension of the operator $\phi^{2}(x)$ in the disordered theory. This would require computing the correlation function $<\sum_{A=1}^{n}\phi_{A}^{2}(x)...>$ in the replicated theory. However, this will not be sufficient since in the replicated theory the operator $\sum_{A=1}^{n}\phi_{A}^{2}(x)$ can mix with the double replicated operator $\sum_{A\neq B=1}^{n}\phi_{A}\phi_{B}(x)$. 
In general, the renormalized correlation function $<\sum_{A=1}^{n}\phi_{A}^{2}(x)\phi_{C}(y)\phi_{D}(z)>$ 
satisfies the Callan-Symanzik equation
\bea\label{Callan-SymanzikPhi2}
&&\Big(\m\frac{\p}{\p\m}+\beta_{\lambda}\frac{\p}{\p\lambda}+\beta_{v}\frac{\p}{\p v}+\g_{\phi^{2}}\Big)<\sum_{A=1}^{n}\phi_{A}^{2}(x)\phi_{C}(y)\phi_{D}(z)>_{\text{rep.}}\nn\\
&&\qquad\qquad\qquad\qquad\qquad\qquad+\g'_{\phi\phi}<\sum_{A\neq B=1}^{n}\phi_{A}\phi_{B}(x)\phi_{C}(y)\phi_{D}(z)>_{\text{rep.}}=0\,.
\eea
To determine the anomalous dimension $\g_{\phi^{2}}$ and the mixing coefficient $\g'_{\phi\phi}$, we compute the 3-point functions $G^{2,1}_{\phi^{2}}=<\sum_{A=1}^{n}\phi_{A}^{2}(x)\phi_{C}(y)\phi_{D}(z)>$ and $G^{2,1}_{\phi\phi}=<\sum_{A\neq B=1}^{n}\phi_{A}\phi_{B}(x)\phi_{C}(y)\phi_{D}(z)>$ up to a 2-loop order. The Feynman diagrams are given in figure \ref{CompositeOperator1}. 
These are given by
\bea
G^{2,1}_{\phi^{2}}&=&\delta_{CD}\Big[2(1+\delta_{\phi^{2}})+\Big(4v(n+2)(1+\delta_{v})(1+\delta_{\phi^{2}})-\lambda(1+\delta_{\lambda})(1+\delta_{\phi^{2}})\Big)I_{1}(\vec p)\nn\\
&&+\Big(\frac{\lambda^{2}}{2}-4v\lambda(n+2)+8v^{2}(n+2)^{2}\Big)I^{2}_{1}(\vec p)+\Big(\lambda^{2}-24v\lambda+48v^{2}(n+2)\Big)I_{3}(\vec p,\vec q)\Big]\,,\nn\\
\eea
and
\bea
G^{2,1}_{\phi\phi}&=&(1-\delta_{CD})\Big[2(1+\delta_{\phi\phi})+8v(1+\delta_{v})(1+\delta_{\phi\phi})I_{1}(\vec p)+32v^{2}I^{2}_{1}(\vec p)\nn\\
&&+\Big(-8v\lambda+16v^{2}(n+6)\Big)I_{3}(\vec p,\vec q)\Big]\,.
\eea
 Requiring that the divergences to cancel, we obtain
 \bea
&&\delta_{\phi^{2}}=\frac{\lambda-4v(n+2)}{4\pi\e}-\frac{(48(n+2)v^{2}-24v\lambda+\lambda^{2})\ln 2}{8\pi^{2}\e}+\frac{8(n+2)(n+5)v^{2}-4(n+5)v\lambda+\lambda^{2}}{8\pi^{2}\e^{2}}\nn\\
&&\delta_{\phi\phi}=-\frac{2v}{\pi\e}+\frac{v(\lambda-2(n+6)v)\ln2}{\pi^{2}\e}+\frac{v(2(n+10)v-\lambda)}{2\pi^{2}\e^{2}}
\eea
Using the above relation and the Callan-Symanzik equation~\eqref{Callan-SymanzikPhi2}, we obtain $\g'_{\phi\phi}=0$ and
\bea
\g_{\phi^{2}}=\frac{\lambda_{*}-4v_{*}(n+2)}{4\pi}-\frac{(48(n+2)v_{*}^{2}-24v_{*}\lambda_{*}+\lambda_{*}^{2})\ln 2}{4\pi^{2}}\,,
\eea
where $\lambda_{*}$ and $v_{*}$ are coupling constants at the fixed point. Thus, the anomalous dimesnion of the $(\text{mass})^{2}$ operator in the disordered theory is
\be
\g_{\phi^{2}}=\frac{\lambda_{*}-8v_{*}}{4\pi}-\frac{(96v_{*}^{2}-24v_{*}\lambda_{*}+\lambda_{*}^{2})\ln 2}{4\pi^{2}}\,.
\ee
\subsection{$\mathcal O(N)$ invariant field theory with a disorder interaction at boundary}
We can generalize the analysis in the previous section to the $\mathcal O(N)$ invariant scalar field theory in $\e$-expansion. In this case, we will find that we have a non-trivial IR disorder fixed point for the values $1<N<4$. The starting action is 
\be
S=\int d^{d}\vec x\,dy\,\frac{1}{2}\p_{\m}\phi^{I}\p^{\m}\phi^{I}+\int d^{d}\vec x\,h(x)\phi^{I}\phi^{I}(x)+\frac{\lambda}{4!}\int d^{d}\vec x\,(\phi^{I}\phi^{I}(x))^{2}\,,
\ee
where $I,J..$ run from $1$ to $N$.
In this case, the replicated action is given by
\be
S_{\text{rep.}}=\sum_{A=1}^{n}S_{A}-\frac{v}{2}\int d^{d}x\,\sum_{A,B=1}^{n}\phi^{2}_{A}\mathcal \phi^{2}_{B}(x)\,,
\ee 
where
\be
S_{A}=\int d^{d}\vec x\,dy\,\frac{1}{2}\p_{\m}\phi_{A}^{I}\p^{\m}\phi_{A}^{I}+\frac{\lambda}{4!}\int d^{d}\vec x\,(\phi_{A}^{I}\phi_{A}^{I}(x))^{2}\,.
\ee
Next, we look for the fixed point of the renormalization group flow in $d=2-\e$-dimensions. We compute the 4-point function $<\phi^{I_{1}}_{A_{1}}\phi^{I_{2}}_{A_{2}}\phi^{I_{3}}_{A_{3}}\phi^{I_{4}}_{A_{4}}>$. The contributions to the 4-point function to 2-loop order are (we have the same Feynman diagrams as shown in the figure~\ref{4-pointFn1})
\bea
&&\delta^{I_{1}I_{2}}\delta^{I_{3}I_{4}}\delta_{A_{1}A_{2}}\delta_{A_{3}A_{4}}\Big(4v(1+\delta_{v})+8v^{2}(1+\delta_{v})^{2}(nN+8)I_{1}(\vec p)\nn\\
&&-\frac{4\lambda v(1+\delta_{\lambda})(1+\delta_{v})}{3}(N+2)I_{1}(\vec p)+16v^{3}(N^{2}n^{2}+6Nn+20)I_{2}(\vec p)\nn\\
&&-4v^{2}\lambda(N^{2}n+2Nn+4N+8)I_{2}(\vec p)+\frac{v\lambda^{2}}{3}(N+2)^{2}I_{2}(\vec p)+64v^{3}(5Nn+22)I_{3}(\vec p,\vec q)\nn\\
&&-64v^{2}\lambda\,(N+2)I_{3}(\vec p,\vec q)+\frac{4v\lambda^{2}}{3}\,(N+2)I_{3}(\vec p,\vec q)\Big)\nn\\
&&+\delta^{I_{1}I_{2}}\delta^{I_{3}I_{4}}\delta_{A_{1}A_{2}A_{3}A_{4}}\Big(-\frac{\lambda(1+\delta_{\lambda})}{3}+\frac{\lambda^{2}(1+\delta_{\lambda})^{2}}{18}(N+8)I_{1}(p)-8\lambda v(1+\delta_{\lambda})(1+\delta_{v})I_{1}(p)\nn\\
&&-48v^{2}\lambda I_{2}(\vec p)+\frac{2v\lambda^{2}}{3}(N+8)I_{2}(\vec p)-\frac{\lambda^{3}}{108}(N^{2}+6N+20)I_{2}(\vec p)\nn\\
&&-16v^{2}\lambda\,(Nn+14)I_{3}(\vec p,\vec q)+\frac{16v\lambda^{2}}{3}\,(N+5)I_{3}(\vec p,\vec q)-\frac{\lambda^{3}}{27}(5N+22)I_{3}(\vec p,\vec q)\Big)\,.
\eea
Requiring the divergences to cancel, we obtain
\bea
&&\delta_{v}=\frac{1}{2\pi\e}\Big(\frac{\lambda}{3}(N+2)-2v(nN+8)\Big)-\frac{1}{12\pi^{2}\e}\Big(48(22+5nN)v^{2}-48v\lambda(2+N)+(2+N)\lambda^{2}\Big)\ln 2\nn\\
&&+\frac{1}{48\pi^{2}\e^{2}}\Big(48(8+nN)^{2}v^{2}-12v\lambda(N+2)(12+nN)+(2+N)(4+N)\lambda^{2}\Big)\nn\\
&&\delta_{\lambda}=\frac{1}{2\pi\e}\Big(\frac{N+8}{6}\lambda-24v\Big)-\frac{1}{36\pi^{2}\e}\Big(432(14+nN)v^{2}-144(5+N)v\lambda+(22+5N)\lambda^{2}\Big)\ln 2\nn\\
&&+\frac{1}{144\pi^{2}\e^{2}}\Big(864(20+nN)v^{2}-72(28+5N)v\lambda+(8+N)^{2}\lambda^{2}\Big)\,.
\eea
Solving the Callan-Symanzik equation
\be
\Big(\m\frac{\p}{\p\m}+\beta_{\lambda}\frac{\p}{\p\lambda}+\beta_{v}\frac{\p}{\p v}\Big)<\phi^{I_{1}}_{A_{1}}\phi^{I_{2}}_{A_{2}}\phi^{I_{3}}_{A_{3}}\phi^{I_{4}}_{A_{4}}>=0,
\ee
we obtain the beta functions of the replicated theory. These are 
\bea
&&\beta_{v}=-\e v-\frac{v\Big(6(8+nN)v-(2+N)\lambda\Big)}{6\pi}-\frac{v\Big(48(22+5nN)v^{2}-48(2+N)v\lambda+(2+N)\lambda^{2}\Big)\ln 2}{6\pi^{2}}\,,\nn\\
&&\beta_{\lambda}=-\e\,\lambda-\frac{\lambda(144v-(8+N)\lambda)}{12\pi}-\frac{\lambda\Big(432(14+nN)v^{2}-144(5+N)v\lambda+(22+5N)\lambda^{2}\Big)\ln2}{18\pi^{2}}\,.\nn\\
\eea
From the above, we obtain the following fixed point in the disordered theory:
\begin{enumerate}
\item {\bf Gaussian fixed point:} $\lambda=0,\,v=0$\,.
\item {\bf Pure fixed point:} $\lambda=\frac{12\pi\e}{N+8}+\frac{96(22+5N)\pi\e^{2}\ln 2}{(N+8)^{3}},\,v=0$\,.
\item {\bf Disorder fixed point:} $\lambda=0,\,v=-\frac{\pi\e}{8}-\frac{11\e^{2}}{32}\pi\ln2$\,.
\item {\bf Mixed disorder fixed point:} $\lambda=\frac{3\pi\e}{N-1}+\frac{3\Big(32+N(15N-128)\Big)\pi\e^{2}\ln 2}{16(N-1)^{3}},\,v=\frac{(4-N)\e\pi}{16(N-1)}+\frac{\Big(128-N(512-(196-55N)N)\Big)\pi\e^{2}\ln2}{256(N-1)^{3}}$\,.
\end{enumerate}
Note that the mixed disorder fixed point is an IR fixed point for the sufficiently small values of $\e$ and $N\in (1,4)$. The fixed point disappears for $N=1$.

Next, we calculate the anomalous dimension of the operator $(\text{mass})^{2}$ in the disordered theory. We compute the anomalous dimension to one loop.
For this, we compute the three point correlation function $<\sum_{A}\phi^{I}_{A}\phi^{I}_{A}(x)\phi^{K}_{B}\phi^{L}_{C}>$ at one loop order. In this case, we get
\bea
&&\delta^{LK}\delta_{BC}\Big[2(1+\delta_{\phi^{2}})+\Big(-\frac{\lambda}{4!}8(N+2)+\frac{v}{2}8(nN+2)\Big)\int\frac{d^{d}k}{(2\pi)^{d}}\frac{1}{|\vec k||\vec k-\vec p|}\Big]\,.
\eea
Cancellation of the divergence requires
\be
\delta_{\phi^{2}}=\frac{\lambda}{6}\frac{N+2}{2\pi\e}-\frac{2v(nN+2)}{2\pi\e}\,.
\ee
Solving the Callan Symanjik equation, we obtain
\be
\g_{\phi^{2}}=\frac{\lambda(N+2)}{12\pi}-\frac{v(nN+2)}{2\pi}\,.
\ee
In the limit $n\rightarrow 0$, the anomalous dimension is
\be
\gamma_{\phi^{2}}=\frac{\lambda(N+2)}{12\pi}-\frac{v}{\pi}\,.
\ee
At the mixed disorder fixed point, the anomalous dimension becomes (for $d=2-\e$)
\be
\Delta_{\phi^{2}}=d-1+\gamma_{\phi^{2}}=1+\frac{(20-11N)\e}{16(N-1)}\,.
\ee
\subsection{Mixed $\sigma\phi$ theory}
\begin{figure}[htpb]
\begin{center}
\vspace{-0.3 cm}
\includegraphics[width=6in]{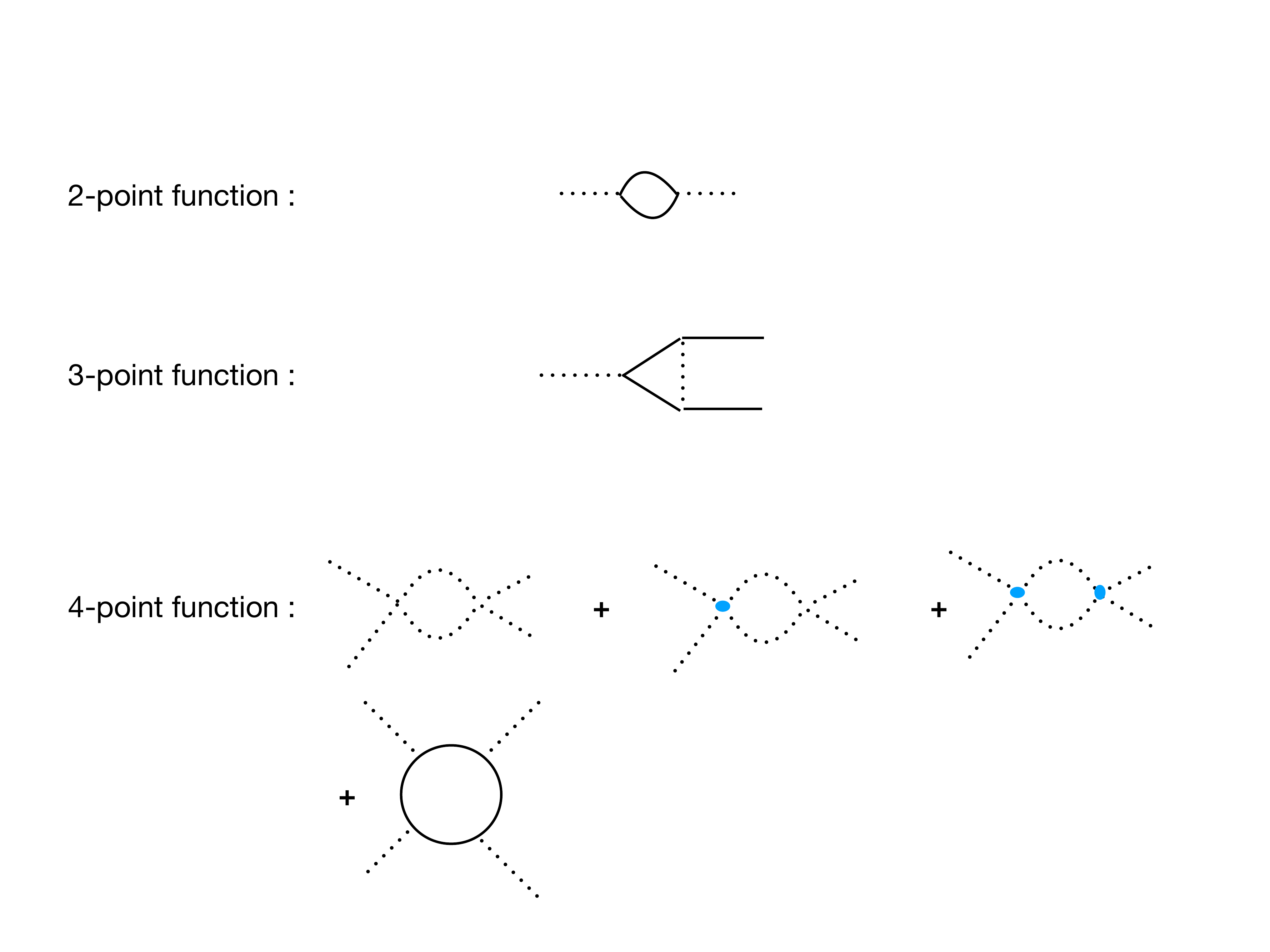}
\vspace{-0.5 cm}
\caption{Boundary $2,3$ and $4$-point function at one loop order. The blue dot represents the disorder vertex. Solid and dotted lines represent the $\phi$-propagator and $\sigma$-propagator, respectively.\label{234-pointFn}}
\end{center}
\end{figure}

Next, we consider a model where the bulk-free scalar fields interact with matter degrees of freedom at the boundary. We focus here on the case where the boundary degrees of freedom consists of a single scalar field $\sigma$. As we will see below, this provides another example of disorder field theory with an IR fixed point with a non zero value of disorder strength when $\e<<1$.
The action is given by,
\be
S=\int d^{d+1}x\,\frac{1}{2}\p_{\m}\phi^{I}\p^{\m}\phi^{I}+\int d^{d}x\,\Big(\frac{1}{2}\p_{\m}\sigma\p^{\m}\sigma+h(x)\sigma^{2}+\frac{\lambda_{1}}{2}\sigma\phi^{I}\phi^{I}+\frac{\lambda_{2}}{4!}\sigma^{4}\Big)\,.
\ee
In the above $I=1,..,N$. Harris criteria dictate that the disorder is marginal in the dimensions $d=4$. We will, therefore, work in the dimensions $d=4-\e$. Note that we have also included boundary interactions which are marginal in the dimensions $d=4$.\\
After averaging over the disorder, we obtain the replicated action given by
\be
S_{\text{repl.}}=\sum_{A=1}^{n}S_{A}-\frac{v}{2}\sum_{A,B=1}^{n}\int d^{d}x\,\sigma_{A}^{2}\sigma_{B}^{2}\,,
\ee
where
\be
S_{A}=\int d^{d+1}x\,\frac{1}{2}\p_{\m}\phi^{I}_{A}\p^{\m}\phi^{I}_{A}+\int d^{d}x\,\Big(\frac{1}{2}\p_{\m}\sigma_{A}\p^{\m}\sigma_{A}+\frac{\lambda_{1}}{2}\sigma_{A}\phi^{I}_{A}\phi^{I}_{A}+\frac{\lambda_{2}}{4!}\sigma_{A}^{4}\Big)\,.
\ee
The replicated theory is a standard local quantum field theory. We will analyze the renormalization flow in the theory and look for the perturbative fixed point. To obtain this, we will compute 2, 3 and 4-point functions involving the scalar field $\sigma$. The relevant Feynman diagrams are shown in figure~\ref{234-pointFn}.
We first begin with the 2-point function of the $\sigma$ field. The 2-point function is divergent at one-loop, and there is a wave function renormalization for the $\sigma$ field. The one loop contribution to the 2-point functions are
\be
<\sigma_{A}(p)\sigma_{B}(-p)>=N\delta_{AB}\frac{\lambda^{2}_{1}}{2}\int\frac{d^{d}\vec k}{(2\pi)^{d}}\frac{1}{|\vec p||\vec k+\vec p|}-\delta_{AB}\,p^{2}\delta_{\sigma}\,.
\ee
The counter term $\delta_{\sigma}$ is determined by requiring that the derivative of the above contribution with respect to $p^{2}$ at some renormalization scale $\m$ is finite.
Cancelling the divergence requires that
\be
\delta_{\sigma}=-\frac{N\lambda^{2}_{1}}{128\pi^{2}\e}\,.
\ee
Next, we look for the 3-point function $<\sigma_{A_{1}}\phi^{I}_{A_{2}}\phi^{J}_{A_{3}}>$ at one loop order. The result is given by
\be
\delta^{IJ}\delta_{A_{1}A_{2}A_{3}}\Big[-\lambda_{1}(1+\delta_{\lambda_{1}})-\lambda^{3}_{1}\int\frac{d^{d}\vec k}{(2\pi)^{d}}\frac{1}{\vec k^{2}|\vec k-\vec q||\vec k+\vec p|}\Big]\,.
\ee
Cancelling the divergence requires that
\be
\delta_{\lambda_{1}}=-\frac{\lambda^{2}_{1}}{8\pi^{2}\e}\,.
\ee 
Finally, we calculate the 4-point function of $\sigma$ field i.e. $<\sigma_{A_{1}}\sigma_{A_{2}}\sigma_{A_{3}}\sigma_{A_{4}}>$. In this case, we get
\bea
&&-\lambda_{2}(1+\delta_{\lambda_{2}})\,\delta_{A_{1}A_{2}A_{3}A_{4}}+4v(1+\delta_{v})\delta_{A_{1}A_{2}}\delta_{A_{3}A_{4}}+\frac{1}{2}\Big[3\lambda_{2}^{2}\delta_{A_{1}A_{2}A_{3}A_{4}}\nn\\
&&-8\lambda_{2}v\Big(\delta_{A_{1}A_{2}}\delta_{A_{3}A_{4}}+6\delta_{A_{1},A_{2},A_{3},A_{4}}\Big)+16v^{2}(n+8)\delta_{A_{1}A_{2}}\delta_{A_{3}A_{4}}\Big]\times\int\frac{d^{d}\vec k}{(2\pi)^{d}}\frac{1}{k^{2}(\vec k+\vec p)^{2}}\nn\\
&&+3N\lambda_{1}^{4}\int \frac{d^{d}\vec k}{(2\pi)^{d}}\frac{1}{|\vec k||\vec k+\vec p||\vec k+\vec p+\vec q||\vec k-\vec r|}\delta_{A_{1}A_{2}A_{3}A_{4}}\,.
\eea
Cancellation of divergences implies that
\be
\lambda_{2}\delta_{\lambda_{2}}=\frac{3\lambda^{2}_{2}-48v\lambda_{2}+6N\lambda^{4}_{1}}{16\pi^{2}\e},\qquad \delta_{v}=\frac{\lambda_{2}-2v(n+8)}{8\pi^{2}\e}\,.
\ee
Thus, we obtain the following $\beta$-function equations
\bea
&&\beta_{1}=-\frac{\e \lambda_{1}}{2}+\frac{(N-32)\lambda^{3}_{1}}{256\pi^{2}}\,,\nn\\
&&\beta_{2}=-\lambda_{2}\e+\frac{3\lambda^{2}_{2}}{16\pi^{2}}+\frac{N\lambda^{2}_{1}\lambda_{2}}{64\pi^{2}}-\frac{3v\lambda_{2}}{\pi^{2}}+\frac{3N\lambda_{1}^{4}}{8\pi^{2}}\,,\nn\\
&&\beta_{v}=-\e v-\frac{(8+n)v^{2}}{4\pi^{2}}+\frac{N\lambda^{2}_{1}v}{64\pi^{2}}+\frac{v\lambda_{2}}{8\pi^{2}}\,.
\eea
The $\beta$-function for the disorderd theory is obtained by substituting $n\rightarrow 0$ limit
\bea
&&\beta_{1}=-\frac{\e \lambda_{1}}{2}+\frac{(N-32)\lambda^{3}_{1}}{256\pi^{2}}\,,\nn\\
&&\beta_{2}=-\lambda_{2}\e+\frac{3\lambda^{2}_{2}}{16\pi^{2}}+\frac{N\lambda^{2}_{1}\lambda_{2}}{64\pi^{2}}-\frac{3v\lambda_{2}}{\pi^{2}}+\frac{3N\lambda_{1}^{4}}{8\pi^{2}}\,,\nn\\
&&\beta_{v}=-\e v-\frac{2v^{2}}{\pi^{2}}+\frac{N\lambda^{2}_{1}v}{64\pi^{2}}+\frac{v\lambda_{2}}{8\pi^{2}}\,.
\eea
Next, we solve the $\beta$-function equation to find the fixed points. We find the following fixed points:
\begin{enumerate}
\item {\bf Gaussian fixed point:} $\lambda_{1}=\lambda_{2}=v=0$\, for any value of $N$.
\item {\bf Decoupled pure fixed point:} $\lambda_{1}=v=0, \,\lambda_{2}=\frac{16\pi^{2}\e}{3}$\,  for any value of $N$.
\item {\bf Coupled pure fixed point:} $v=0,\,\lambda^{2}_{1}=\frac{128\pi^{2}\e}{N-32},\,\\\lambda_{2}=\frac{8\pi^{2}\e}{3(N-32)}\Big(-(N+32)+\sqrt{1024+N(N-4544)}\Big)$. The fixed point exists for $N>4544$. However, the coupling constant $\lambda_{2}$ is negative and, therefore, the fixed point is nonperturbatively unstable. 
\item {\bf Coupled mixed disorder fixed point:} $\lambda^{2}_{1}=\frac{128\pi^{2}\e}{N-32},\,v=\frac{(1024+1600N+N^{2})\pi^{2}\e}{2(N-32)(N+32)},\,\lambda_{2}=\frac{12288N\pi^{2}\e}{(N-32)(N+32)}$. The fixed point is non-perturbatively stable as long as $N>32$. \\

We can also compute the anomalous dimension of the field $\sigma$ at the fixed point at one-loop order. It is given by
\be
\g_{\sigma}=\frac{\m}{2(1+\delta_{\sigma})}\frac{d}{d\m}(1+\delta_{\sigma})=\frac{N\lambda_{1}^{2}}{256\pi^{2}}=\frac{N\e}{2(N-32)}\,.
\ee
\end{enumerate}
\section{Anomalous dimensions of displacement operators at one loop}
This section discusses the effect of classical disorder on the one-loop computation of anomalous dimensions of higher spin displacement operators. It was shown in~\cite{Giombi:2019enr} that in the pure theory with $\phi^{4}$ interaction at the boundary, describe by the action~\eqref{SingleScalar} with vanishing disorder coupling, the higher spin displacement operators do not have anomalous dimensions at two-loop order. Here, we repeat their analysis in the presence of classical disorder and find that the anomalous dimension is not affected, i.e. the displacement operators do not have anomalous dimensions. 

We will compute the anomalous dimension of the displacement operator at one loop for a single scalar field with action given in~\eqref{SingleScalar}. The corresponding replicated action is given in~\eqref{ReplicatedSingleScalar}. The energy-momentum tensor of the replicated theory is 
\be
T^{\text{Repl.}}_{\m\n}=\sum_{A}T_{\m\n}^{A}\,,
\ee
where
\be
T^{A}_{\m\n}=\p_{\m}\phi_{A}\p_{\n}\phi_{A}-\frac{\delta_{\m\n}}{2}\p_{\rho}\phi_{A}\p^{\rho}\phi_{A}-\frac{d-1}{4d}(\p_{\m}\p_{\n}-\delta_{\m\n}\p^{2})\phi_{A}^{2}\,,
\ee 
Note that we have considered only the bulk contributions to the energy-momentum tensor. The boundary terms in the energy-momentum tensor are proportional to the coupling constant and, therefore, in the anomalous dimension computation at one loop, these will be higher-order contributions in the powers of the coupling constant. 
The displacement operator of the replicated theory is
\be
D^{\text{Repl.}}(\vec x)=\lim_{y\rightarrow 0}T^{\text{Repl.}}_{yy}\,.
\ee 
Following the calculation of~\cite{Giombi:2019enr}, we see that at one loop we do not have any divergences for $d=2$.  
In particular, at one loop the contribution is proportional to the integral
\be
\int\frac{d^{d}\vec k}{(2\pi)^{d}}\frac{1}{|\vec k||\vec k+\vec p|}\Big(\frac{d}{2}(\vec k^{2}+(\vec k+\vec p)^{2})-\frac{\vec p^{2}}{2}\Big)
\ee
which evaluates to zero for $d=2$. Thus, at one-loop, we do not have anomalous dimensions. The calculation proceeds similarly for the higher spin displacement operators. Basically, up to an overall factor, which depends on the coupling constants and some numerical factors, the one loop integrals are the same as in~\cite{Giombi:2019enr}. In the paper, authors have shown that those integrals, appearing at one-loop order, are finite for $d=2$ and thus, there are no divergences. We conclude, therefore, that the disorder does not have any effect on the scaling dimension. 
\section{Quantum disorder}
Finally, we consider an example of a scalar field theory with the quantum disorder localized at the boundary. As we have explained before, there is an extra time coordinate and the disorder field $h(\vec x)$ is homogeneous in time. We consider the theory of a single scalar field with action
\be
S=\int d^{d}\vec x\,dt\,dy\,\frac{1}{2}\p_{\m}\phi\p^{\m}\phi+\int d^{d}\vec x\,dt\,h(\vec x)\phi^{2}(\vec x,t)\,.
\ee
For $d=2$, the disorder coupling at the boundary is marginal. The only other operator which is marginal in $d=2$ is $\phi^{3}$. However, we will assume the coupling constant of the $\phi^{3}$ term to be zero.\\
After averaging over the disorder, we obtain the replicated action, which is
\be
S_{\text{repl.}}=\sum_{A=1}^{n}\int d^{d}\vec x\,dt\,dy\,\frac{1}{2}\p_{\m}\phi_{A}\p^{\m}\phi_{A}-\frac{v}{2}\sum_{A,B=1}^{n}\int d^{d}x\,dt\,dt'\phi_{A}\phi_{A}(\vec x,t)\phi_{B}\phi_{B}(\vec x,t')\,.
\ee 
Note that it is not a standard local quantum field theory. However, we can analyze the theory in powers of disorder strength in perturbation expansion. 
To find the fixed point of the theory, we compute the correlation function of boundary operators\\
$<\phi_{A_{1}}(\vec p_{1},E_{1})\phi_{A_{2}}(\vec p_{2},E_{2})\phi_{A_{3}}(\vec p_{3},E_{3})\phi_{A_{4}}(\vec p_{4},E_{4})>$. It is given by (upto an overall factor which enforces conservation of momenta)
\bea
&&8\frac{v(1+\delta_{v})}{2}\delta_{A_{1}A_{2}}\delta_{A_{3}A_{4}}\delta(E_{1}+E_{2})\delta(E_{3}+E_{4})+128\frac{v^{2}}{8}\delta_{A_{1}A_{2}}\delta_{A_{3}A_{4}}\delta(E_{1}+E_{2})\delta(E_{3}+E_{4})\nn\\
&&\int \frac{d^{d}\vec k}{(2\pi)^{d}}\Big[\frac{1}{\sqrt{(\vec k^{2}+E_{3}^{2})((\vec p-\vec k)^{2}+E_{3}^{2})}}+\frac{1}{\sqrt{(\vec k^{2}+E_{1}^{2})((\vec p-\vec k)^{2}+E_{1}^{2})}}\nn\\
&&+\frac{2}{\sqrt{(\vec k^{2}+E_{1}^{2})((\vec p\,'-\vec k)^{2}+E_{3}^{2})}}\Big]\,.
\eea
In the above $\vec p=\vec p_{1}+\vec p_{2}$ and $\vec p\,'=\vec p_{1}+\vec p_{3}$.
Note that in the above, we have excluded the infrared divergent contribution.
Now, in the dimensions, $d=2+\e$, each of above the integral is UV divergent and the divergence is given by
\be
\int\frac{d^{d}\vec k}{(2\pi)^{d}}\frac{1}{\sqrt{(\vec k^{2}+E_{3}^{2})((\vec p-\vec k)^{2}+E_{3}^{2})}}\sim-\frac{1}{2\pi\e}\,.
\ee
Cancellation of divergence requires that
\be
\delta_{v}=\frac{8v}{\pi\e}\,.
\ee
Solving the Callan-Symanizik equation, we obtain the beta function
\be
\beta_{v}=\e v-\frac{8v^{2}}{\pi}\,.
\ee
Thus, it has a UV fixed point given by
\be
v_{*}=\frac{\pi\e}{8}\,.
\ee
From the Callan-Symanizik equation, we can compute the dynamical exponent $\g_{t}$ at the fixed point. At the one-loop order in perturbation theory, we find that  $\g^{*}_{t}$ vanishes, i.e. $\g^{*}_{t}=\mathcal O(v_{*}^{2})$. Thus, the correlation function at one-loop order does not have Lifshitz scaling.

\section{Discussion}
In this article, we study the renormalization group property of a disordered quantum field theory in the presence of a boundary. We constructed examples of boundary field theory with both classical and quantum disorder localized at the boundary. In these theories, we found fixed points of renormalization group flow and computed the anomalous dimension of certain operators. 

There are many interesting questions and directions which have not been addressed in the paper. One of the questions which will be interesting to understand is whether the disorder fixed point, if it exists, exhibits (or under what conditions) conformal symmetry. It will be very interesting since it will lead to examples of disorder conformal field theory, and 
one can use various tools of conformal field theory, see for example~\cite{Komargodski:2016auf} to study the disordered theory at the fixed point.

In another direction, it will be interesting to find an analogous monotonicity theorem that exists in boundary quantum field theory~\cite{PhysRevLett.67.161,Jensen:2015swa, Casini:2016fgb, Nozaki:2012qd, Wang:2021mdq, Kobayashi:2018lil}. It requires finding a quantitative measure that decreases along with the renormalization group flow in a disordered theory.

In the present article, we have studied quench disorder in scalar field theories. It will also be interesting to consider examples with fermions and gauge fields. In this direction, the most interesting model to study would be the mixed dimensional QED and possibly generalization to non-abelian theories. 
\section*{Acknowledgments}
 We thank Justin R. David and Meenu for useful discussion. This work is supported by the ISIRD grant 9-406/2019/IITRPR/5480.
\appendix
\section{Some useful integration}\label{UsefulIntegrations}
In the main text, we encounter integrals of the form 
\be
\int \frac{d^{d}q}{(2\pi)^{d}}\frac{1}{q^{2\alpha}(p+q)^{2\beta}}\,.
\ee
We can evaluate the above integral using the Feynman parametrization
\bea
\frac{1}{q^{2\alpha}(p+q)^{2\beta}}&=&\frac{\Gamma(\alpha+\beta)}{\Gamma(\alpha)\Gamma(\beta)}\int^{1}_{0} dx\,\frac{x^{\alpha-1}(1-x)^{\beta-1}}{(xq^{2}+(1-x)(p+q)^{2})^{\alpha+\beta}}\,,\nn\\
&=&\frac{\Gamma(\alpha+\beta)}{\Gamma(\alpha)\Gamma(\beta)}\int^{1}_{0} dx\,\frac{x^{\alpha-1}(1-x)^{\beta-1}}{(\ell^{2}+x(1-x)p^{2})^{\alpha+\beta}}\,,
\eea
where $\ell^{\m}=q^{\m}+(1-x)p^{\m}$.
The final integral can be evaluated as follows:
\bea
\int \frac{d^{d}q}{(2\pi)^{d}}\frac{1}{q^{2\alpha}(p+q)^{2\beta}}&=&\int \frac{d^{d}q}{(2\pi)^{d}}\frac{\Gamma(\alpha+\beta)}{\Gamma(\alpha)\Gamma(\beta)}\int^{1}_{0} dx\frac{x^{\alpha-1}(1-x)^{\beta-1}}{(q^{2}+x(1-x)p^{2})^{\alpha+\beta}}\nn\\
&=&\int \frac{d^{d}q}{(2\pi)^{d}}\frac{1}{\Gamma(\alpha)\Gamma(\beta)}\int^{1}_{0} dx\,x^{\alpha-1}(1-x)^{\beta-1}\,\int_{0}^{\infty}dt\,t^{\alpha+\beta-1}\,e^{-t(q^{2}+x(1-x)p^{2})}\nn\\
&=&\frac{V_{d-1}\Gamma(\frac{d}{2})}{2(2\pi)^{d}\Gamma(\alpha)\Gamma(\beta)}\int^{1}_{0} dx\,x^{\alpha-1}(1-x)^{\beta-1}\,\int_{0}^{\infty}dt\,t^{\alpha+\beta-1-\frac{d}{2}}\,e^{-tx(1-x)p^{2}}\nn\\
&=&\frac{V_{d-1}\Gamma(\frac{d}{2})}{2(2\pi)^{d}\Gamma(\alpha)\Gamma(\beta)}\int^{1}_{0} dx\,x^{\alpha-1}(1-x)^{\beta-1}\,\frac{\Gamma(\alpha+\beta-\frac{d}{2})}{(x(1-x)p^{2})^{\alpha+\beta-\frac{d}{2}}}\nn\\
&=&\frac{V_{d-1}\Gamma(\frac{d}{2})\Gamma(\alpha+\beta-\frac{d}{2})}{2(2\pi)^{d}\Gamma(\alpha)\Gamma(\beta)(p^{2})^{\alpha+\beta-\frac{d}{2}}}\int^{1}_{0} dx\,x^{\frac{d}{2}-\beta-1}(1-x)^{\frac{d}{2}-\alpha-1}\,\nn\\
&=&\frac{V_{d-1}\Gamma(\frac{d}{2})\Gamma(\alpha+\beta-\frac{d}{2})}{2(2\pi)^{d}\Gamma(\alpha)\Gamma(\beta)(p^{2})^{\alpha+\beta-\frac{d}{2}}}\frac{\Gamma(\frac{d}{2}-\beta)\Gamma(\frac{d}{2}-\alpha)}{\Gamma(d-\alpha-\beta)}\nn\\
&=&\frac{\Gamma(\alpha+\beta-\frac{d}{2})}{(4\pi)^{\frac{d}{2}}\Gamma(\alpha)\Gamma(\beta)(p^{2})^{\alpha+\beta-\frac{d}{2}}}\frac{\Gamma(\frac{d}{2}-\beta)\Gamma(\frac{d}{2}-\alpha)}{\Gamma(d-\alpha-\beta)}
\eea
Using the above integration formula, some of the integrals used in the main text are given below.
\bea
&&I_{1}(\vec p)=\int\frac{d^{d}\vec k}{(2\pi)^{d}}\frac{1}{|\vec k||\vec k+\vec p|}=\frac{\Gamma(\frac{d-1}{2})^{2}\Gamma(1-\frac{d}{2})}{(4\pi)^{\frac{d}{2}}\pi (p^{2})^{1-\frac{d}{2}}\Gamma(d-1)}\,,\nn\\
&& I_{2}(\vec p)=\int\frac{d^{d}\vec k_{1}\,d^{d}\vec k_{2}}{(2\pi)^{2d}}\frac{1}{|\vec k_{1}+\vec p||\vec k_{1}||\vec k_{2}+\vec p||\vec k_{2}|}=I_{1}(\vec p)^{2}\,,\nn\\
&&I_{3}(\vec p,\vec q)=\int\frac{d^{d}\vec k_{1}\,d^{d}\vec k_{2}}{(2\pi)^{2d}}\frac{1}{|\vec k_{1}+\vec p||\vec k_{1}||\vec k_{1}-\vec k_{2}-\vec q||\vec k_{2}|}\Big|_{\vec q=0}=\frac{\Gamma(1-\frac{d}{2})\Gamma(\frac{d-1}{2})^{3}\Gamma(2-d)\Gamma(d-\frac{3}{2})}{(4\pi)^{d}\pi^{\frac{3}{2}}\Gamma(\frac{3-d}{2})(p^{2})^{2-d}\Gamma(d-1)\Gamma(\frac{3d}{2}-2)}\,.\nn\\
\eea


\begin{thebibliography}{10}

\bibitem{Harris_1974}
A.~B. Harris, {\it Effect of random defects on the critical behaviour of ising
  models},  {\em Journal of Physics C: Solid State Physics} {\bf 7} (may, 1974)
  1671--1692.

\bibitem{PhysRevB.26.154}
D.~Boyanovsky and J.~L. Cardy, {\it Critical behavior of $m$-component magnets
  with correlated impurities},  {\em Phys. Rev. B} {\bf 26} (Jul, 1982)
  154--170.

\bibitem{Dotsenko:1994im}
V.~Dotsenko, M.~Picco, and P.~Pujol, {\it {Spin spin critical point correlation
  functions for the 2-D random bond Ising and Potts models}},  {\em Phys. Lett.
  B} {\bf 347} (1995) 113--119,
  [\href{http://xxx.lanl.gov/abs/hep-th/9405003}{{\tt hep-th/9405003}}].

\bibitem{Dotsenko:1994sy}
V.~Dotsenko, M.~Picco, and P.~Pujol, {\it {Renormalization group calculation of
  correlation functions for the 2-d random bond Ising and Potts models}},  {\em
  Nucl. Phys. B} {\bf 455} (1995) 701--723,
  [\href{http://xxx.lanl.gov/abs/hep-th/9501017}{{\tt hep-th/9501017}}].

\bibitem{Fujita:2008rs}
M.~Fujita, Y.~Hikida, S.~Ryu, and T.~Takayanagi, {\it {Disordered Systems and
  the Replica Method in AdS/CFT}},  {\em JHEP} {\bf 12} (2008) 065,
  [\href{http://xxx.lanl.gov/abs/0810.5394}{{\tt arXiv:0810.5394}}].

\bibitem{Hartnoll:2014cua}
S.~A. Hartnoll and J.~E. Santos, {\it {Disordered horizons: Holography of
  randomly disordered fixed points}},  {\em Phys. Rev. Lett.} {\bf 112} (2014)
  231601, [\href{http://xxx.lanl.gov/abs/1402.0872}{{\tt arXiv:1402.0872}}].

\bibitem{Hartnoll:2015rza}
S.~A. Hartnoll, D.~M. Ramirez, and J.~E. Santos, {\it {Thermal conductivity at
  a disordered quantum critical point}},  {\em JHEP} {\bf 04} (2016) 022,
  [\href{http://xxx.lanl.gov/abs/1508.04435}{{\tt arXiv:1508.04435}}].

\bibitem{Aharony:2015aea}
O.~Aharony, Z.~Komargodski, and S.~Yankielowicz, {\it {Disorder in Large-N
  Theories}},  {\em JHEP} {\bf 04} (2016) 013,
  [\href{http://xxx.lanl.gov/abs/1509.02547}{{\tt arXiv:1509.02547}}].

\bibitem{Aharony:2018mjm}
O.~Aharony and V.~Narovlansky, {\it {Renormalization group flow in field
  theories with quenched disorder}},  {\em Phys. Rev. D} {\bf 98} (2018), no.~4
  045012, [\href{http://xxx.lanl.gov/abs/1803.08534}{{\tt arXiv:1803.08534}}].

\bibitem{Narovlansky:2018muj}
V.~Narovlansky and O.~Aharony, {\it {Renormalization Group in Field Theories
  with Quantum Quenched Disorder}},  {\em Phys. Rev. Lett.} {\bf 121} (2018),
  no.~7 071601, [\href{http://xxx.lanl.gov/abs/1803.08529}{{\tt
  arXiv:1803.08529}}].

\bibitem{Callan:1993mw}
C.~G. Callan, Jr. and I.~R. Klebanov, {\it {Exact C = 1 boundary conformal
  field theories}},  {\em Phys. Rev. Lett.} {\bf 72} (1994) 1968--1971,
  [\href{http://xxx.lanl.gov/abs/hep-th/9311092}{{\tt hep-th/9311092}}].

\bibitem{Callan:1994ub}
C.~G. Callan, I.~R. Klebanov, A.~W.~W. Ludwig, and J.~M. Maldacena, {\it {Exact
  solution of a boundary conformal field theory}},  {\em Nucl. Phys. B} {\bf
  422} (1994) 417--448, [\href{http://xxx.lanl.gov/abs/hep-th/9402113}{{\tt
  hep-th/9402113}}].

\bibitem{Fendley:1994rh}
P.~Fendley, H.~Saleur, and N.~P. Warner, {\it {Exact solution of a massless
  scalar field with a relevant boundary interaction}},  {\em Nucl. Phys. B}
  {\bf 430} (1994) 577--596,
  [\href{http://xxx.lanl.gov/abs/hep-th/9406125}{{\tt hep-th/9406125}}].

\bibitem{Herzog:2017xha}
C.~P. Herzog and K.-W. Huang, {\it {Boundary Conformal Field Theory and a
  Boundary Central Charge}},  {\em JHEP} {\bf 10} (2017) 189,
  [\href{http://xxx.lanl.gov/abs/1707.06224}{{\tt arXiv:1707.06224}}].

\bibitem{Giombi:2019enr}
S.~Giombi and H.~Khanchandani, {\it {$O(N)$ models with boundary interactions
  and their long range generalizations}},  {\em JHEP} {\bf 08} (2020), no.~08
  010, [\href{http://xxx.lanl.gov/abs/1912.08169}{{\tt arXiv:1912.08169}}].

\bibitem{Giombi:2016ejx}
S.~Giombi, {\it {Higher Spin \textemdash{} CFT Duality}},  in {\em {Theoretical
  Advanced Study Institute in Elementary Particle Physics}: {New Frontiers in
  Fields and Strings}}, 7, 2016.
\newblock \href{http://xxx.lanl.gov/abs/1607.02967}{{\tt arXiv:1607.02967}}.

\bibitem{Komargodski:2016auf}
Z.~Komargodski and D.~Simmons-Duffin, {\it {The Random-Bond Ising Model in 2.01
  and 3 Dimensions}},  {\em J. Phys. A} {\bf 50} (2017), no.~15 154001,
  [\href{http://xxx.lanl.gov/abs/1603.04444}{{\tt arXiv:1603.04444}}].

\bibitem{PhysRevLett.67.161}
I.~Affleck and A.~W.~W. Ludwig, {\it Universal noninteger ``ground-state
  degeneracy'' in critical quantum systems},  {\em Phys. Rev. Lett.} {\bf 67}
  (Jul, 1991) 161--164.

\bibitem{Jensen:2015swa}
K.~Jensen and A.~O'Bannon, {\it {Constraint on Defect and Boundary
  Renormalization Group Flows}},  {\em Phys. Rev. Lett.} {\bf 116} (2016),
  no.~9 091601, [\href{http://xxx.lanl.gov/abs/1509.02160}{{\tt
  arXiv:1509.02160}}].

\bibitem{Casini:2016fgb}
H.~Casini, I.~Salazar~Landea, and G.~Torroba, {\it {The g-theorem and quantum
  information theory}},  {\em JHEP} {\bf 10} (2016) 140,
  [\href{http://xxx.lanl.gov/abs/1607.00390}{{\tt arXiv:1607.00390}}].

\bibitem{Nozaki:2012qd}
M.~Nozaki, T.~Takayanagi, and T.~Ugajin, {\it {Central Charges for BCFTs and
  Holography}},  {\em JHEP} {\bf 06} (2012) 066,
  [\href{http://xxx.lanl.gov/abs/1205.1573}{{\tt arXiv:1205.1573}}].

\bibitem{Wang:2021mdq}
Y.~Wang, {\it {Defect $a$-Theorem and $a$-Maximization}},
  \href{http://xxx.lanl.gov/abs/2101.12648}{{\tt arXiv:2101.12648}}.

\bibitem{Kobayashi:2018lil}
N.~Kobayashi, T.~Nishioka, Y.~Sato, and K.~Watanabe, {\it {Towards a
  $C$-theorem in defect CFT}},  {\em JHEP} {\bf 01} (2019) 039,
  [\href{http://xxx.lanl.gov/abs/1810.06995}{{\tt arXiv:1810.06995}}].

\end{thebibliography}


\end{document}